\documentstyle[multicol,aps]{revtex}
\begin{document}
\draft

\title{Electronic structure and tunneling resonance spectra of nanoscopic
aluminum islands}

\author{Gustavo A. Narvaez and George Kirczenow}

\address{Department of Physics, Simon Fraser University, Burnaby,
British Columbia, Canada, V5A 1S6}

\date{\today}
\maketitle

\begin{abstract}

The electronic structure of nanoscopic oxide-coated aluminum islands
is investigated using a tight-binding model that
incorporates the geometry, chemistry and
disorder of the particle. The oxide coat is found to
significantly increase the volume accessible to electrons at the Fermi level.
The level statistics agree with
random matrix theory predictions. States near the Fermi
level show pronounced clustering regardless of disorder. It is
suggested that the observed clusters of tunneling resonances may have a more
complex origin than if they were solely due to many-body non-equilibrium
effects.
\end{abstract}

\begin{multicols}{2}

%
%
%
It is expected that the statistics of the discrete
energy spectra of small disordered metal grains should be
described by random matrix theory (RMT)\cite{RMT}.
Experimental confirmation of RMT predictions for such
systems, however, presents a challenge. Recently, some of
the technological difficulties have been overcome and
electron tunneling spectroscopy experiments have been
carried out on {\em Al}, {\em Au}   and {\em Co}
nanoscopic
particles\cite{ralph_experiments,davidovic_PRL_1999,gueron_PRL_1999}.
These transport experiments  have  opened the possibility of
testing directly the applicability of RMT to the energy spectra of these
nanoscopic metal islands (nanoislands).

The observed electron
tunneling spectra of {\em Al} nanoislands coated
with aluminum oxide\cite{ralph_experiments} showed a surprisingly high
density and pronounced clustering of conductance resonances. Agam and
co-workers
\cite{agam_PRL_1999} proposed that this unexpected behavior is a
manifestation of the effects of electron-electron interactions on
non-equilibrium states of the nanoislands. By focusing on the many-body
aspects of the problem, they were able to account for the anomalous
features of the data qualitatively in an appealing way. However, they
modeled the underlying one-electron spectra of the nanoislands
phenomenologically. For a complete
understanding of the experiments a microscopic treatment of the
single-electron energy spectra is desirable. Microscopic
models of aluminum nano-structures
have been proposed\cite{microscopic_models}, but none of
those works addressed the electronic structure or transport
properties of {\em Al} nanoislands similar to those in the above
experiments.
Recently, Campbell {\em et al.} performed state-of-the-art molecular
dynamics simulations of the passivation of {\em Al}
nanospheres with aluminum oxide but did not calculate
the electronic spectra of the particles\cite{campbell_PRL_1999}.

In this communication, we present microscopic tight-binding
calculations of the
single-electron electronic spectra of {\em Al}
nanoislands coated with {\em Al}-oxide. We include the geometry,
atomic lattice, chemistry and disorder of the nanoparticle
explicitly in our theory.
Our {\em leitmotiv} is twofold: i)
comparison of RMT predictions for the energy spectra of generic
disordered metal grains with the electronic structure obtained from
realistic numerical simulations; and ii)  improvement of our understanding of
the experimentally observed tunneling resonance spectra.

Motivated by the experiment, we consider a hemispheric nanoparticle
of volume ${\cal V}
\sim 40 nm^3$, and investigate the effects of the
disorder at the metal/oxide interface on the
electronic structure and transport properties. (Different
geometries yield qualitatively similar results.) We
start with an analysis of the mean level spacing and
level fluctuations  in the
neighborhood of the Fermi energy of the particles. We find
that describing the discrete energy
levels in this region according to RMT should be a
satisfactory starting point when dealing with {\em Al} nanoislands
coated with {\em Al}-oxide. However, the magnitude of the
calculated mean level spacing is consistent with a significant enhancement
of the effective electronic volume of metallic  nanoparticle due to the
presence of the oxide coat.
Next, a closer look at a few levels around
the Fermi level reveals a rich electronic structure: There are clusters of
levels and the clustering is strongly affected by the specifics
of disorder. Furthermore, the clustering of
the single-electron energy levels persists even in the presence of severe
interface disorder.
Finally, we model the differential conductance ($dI/dV$) spectrum and
find clusters of tunneling resonances. The clustering is
due to the complex structure of energy levels around the Fermi
energy of the particle.
   The clusters  of resonances are sensitive to disorder and to the
capacitances of the
tunnel junctions.
The predicted number of tunneling
resonances is not sufficient to explain the experimental
data\cite{ralph_experiments} within a purely single-electron picture of
transport. Nonetheless, our  results suggest that the observed clusters of
tunneling resonances \cite{ralph_experiments} may have a more complex origin
than if they were solely due to many-body non-equilibrium effects during
transport.

%

%
%
%
%
We assume  that the {\em Al} atoms in
the interior of the nanoparticle form an fcc lattice with a lattice
parameter $a=0.405nm$.
   %
	%
The atoms at the surface of the nanoparticle
have a different coordination number from the bulk fcc lattice. This
establishes a criterion for defining the surface of the particle. In
addition, the {\em Al} nanoislands in the experiment  were
passivated with a thin {\em Al}-oxide layer which acts as
a tunnel barrier between the particle and leads.
This layer also imposes some structural disorder at the metal-oxide
interface.  We model this metal-oxide interface in the following way:
First, we define which atoms of the particle belong to the surface and
identify this region as the interface; second,  we
   randomly  choose some of those sites (approximately 50\%) to represent
{\em O} atoms while the others correspond to {\em Al}. In the first part of
this procedure we adopt two different  criteria: Criterion 1
(${\tt C_1}$): We assume the surface atoms to be those whose first neighbor
coordination number is not equal to that of an fcc lattice. Criterion
2 (${\tt C_2}$): Surface atoms are those whose first and/or second neighbor
coordination numbers are not equal to those of an fcc lattice. The random
choice of {\em O} and {\em Al} atoms in the metal-oxide interface is the
only source of disorder in the results presented here.


We use a multiband ({\em s}, {\em p}, and {\em d}
valence orbitals) tight-binding Hamiltonian without spin-orbit coupling:

\begin{equation}
{\cal
H}=\sum_{i,\alpha}{\varepsilon}_{i}^{\alpha}c^{\dagger}_{i\alpha}c_{i\alpha}+
\sum_{i,j,\alpha,\beta}
{\tau}_{{j\alpha},{i\beta}}\left(c^{\dagger}_{i\beta}c_{j\alpha}
+c^{\dagger}_{j\alpha}c_{i\beta}\right)
\label{tb_hamiltonian}
\end{equation}
\noindent $i,j$ label the atoms,
$c^{\dagger}_{i\alpha}(c_{i\alpha})$ creates(destroys) an
electron on site $i$, and the index
$\alpha(\beta)$ indicates the {\em s}, {\em p} or {\em d} orbital.
${\varepsilon}_{i}^{\alpha}$ and ${\tau}_{{j\alpha},{i\beta}}$
are the Slater-Koster (SK) \cite{slater_PR_1954} on-site and hopping
(up to second
neighbors) parameters. We adopt an SK model in which the atomic
orbitals on different sites are orthonormal.
   For the {\em Al}
sites in the interior (bulk) of the particle we used the  SK
parameters  given in Ref. \onlinecite{papaconstantopoulos_book_1986}.
The {\em Al} and {\em O} sites in the metal-oxide
interface  have different SK parameters as is discussed below.
The bonding of the atoms that form the oxide layer is due
to charge transfer from the {\em Al} to the {\em O}.
%
%
Campbell {\em et al.}   have shown that the
metal-oxide interface of spherical {\em Al} nanoparticles consists mainly
of intercalated
$O^{-1/2}$ and $Al^{+1/2}$\cite{campbell_PRL_1999}.
These valence charges  determine the SK
on-site parameters for the interface atoms
which we calculate following the Mulliken-Wolfsberg-Helmholz
molecular-orbital approach\cite{mcglynn_book}. We determine the nearest
neighbor {\em O-Al} hopping parameters by assuming  an average separation
of $0.18nm$\cite{campbell_PRL_1999} between {\em O} and {\em Al} atoms
in the oxide, applying Harrison's model to
obtain the two-center transfer integrals
\cite{Harrison_book_1980} and then transforming them to find the SK
hopping parameters\cite{papaconstantopoulos_book_1986}.
The same procedure is applied to find the nearest neighbor
{\em O-O} hopping, using $0.30nm$\cite{ansell_PRL_1997} as the average
separation between {\em O} atoms. The SK hopping parameters
between {\em Al} atoms are assumed to be the same regardless of
the valence charge.

%
%

   We now present our results for a
nanoisland of volume  $ {\cal V} = 40.45 nm^3 $ for different realizations of
disorder and surface criteria ${\tt C_1}$ and
${\tt C_2}$.   For this volume the
particle-in-a-box (SP) mean level spacing
$\delta^{Al}_{SP}=(4 E^{Al}_F/3 {\cal N}){\cal V}^{-1}= 2.09 meV$
where $E^{Al}_F$ is
the Fermi energy and ${\cal N}$ the electron density of bulk
{\em Al}.
The {\em total} number of atoms   is $N = 2587$ and the
surface-to-bulk ratio is $(N_S/N_B)_1 = 0.538$ and $(N_S/N_B)_2 = 0.970$ for
criteria
${\tt C_1}$ and ${\tt C_2}$, respectively.
Note that
for ${\tt C_1}$ the nanoisland is primarily {\em bulk} while for
${\tt C_2}$ it is almost evenly balanced between {\em bulk} and
    {\em surface}. In the latter case the disorder is
particularly severe. In Table \ref{Table_1} we present, for both ${\tt
C_1}$ and
${\tt C_2}$, the number of $O^{-1/2}$ ($N_{O}$) and $Al^{+1/2}$
($N_{Al}$) atoms, the total number of electrons ($n$) in the nanoisland,
the calculated Fermi energy ($E_F$), and the calculated mean energy level
separation
$(\delta)$. $E_F$ is the energy of the  highest
occupied level assuming that spin up and down
levels are degenerate. We find that the value of $E_F$ does not show
major changes when switching from
${\tt C_1}$ to ${\tt C_2}$. It seems that the increase in  {\em n},  due to
the increased metal-oxide interface, in going from ${\tt C_1}$ to
${\tt C_2}$ is compensated by the associated changes in the
electronic structure of the particles.
This compensation keeps $E_F$ almost independent of
the surface criteria.
Surprisingly, we see  that  the different
choices of surface criteria  have a  minimal effect on  the
values of $\delta$.
Additionally, these values  of $\delta$ agree well with
$\delta^{Al}_{SP}$. These two findings are non-trivial and important:
They imply that the mean level separation around $E_F$
is determined not by the volume of the aluminum core of
the nanoparticle but by a larger volume that includes part of
the oxide coat. For small particles with large surface
to volume ratios this should have a significant effect (exceeding a
factor of 2) on the mean level spacing which should be reflected in the
experimental spectra.  However,
a qualitatively similar enhancement of the value of the nanoisland volume
that is deduced from the experimental capacitance measurements
of the tunnel junctions\cite{ralph_experiments} may also be possible.
In this work the
total volume
${\cal V}$ that determines $\delta^{Al}_{SP}$ is kept fixed at an
experimentally estimated value for the different disorder realizations and
surface criteria.

%
%


We now discuss the statistics of the
energy levels that we find around $E_F$. In each spectrum, we
select a symmetric energy interval (of $\sim 400meV$)
around $E_F$ containing $p=200$ levels.
Then we find the nearest level
energy spacing $E_t=\varepsilon_{t+1}-\varepsilon_{t}$ for each
level $|t\rangle$ in the chosen set, and the mean
level spacing $\delta= (1/p)\sum^{p}_{t=1} E_{t}$.
To compare our results with the RMT prediction, we build a histogram of
the calculated spacings using a bin size $\Delta =\delta/5$.
Figure \ref{Fig_1}(a) shows such histograms for representative
realizations of criteria ${\tt C_1}$ and ${\tt C_2}$. The energy
is in normalized units: $s=E_t / \delta$.  In RMT
our system falls into the   orthogonal ensemble due to its
time reversal symmetry. In this case, the Wigner distribution $P_W(s)=(\pi
s/2)exp(-\pi s^2/4)$ is a good approximation to the
distribution predicted for the Gaussian orthogonal ensemble
(GOE) for the probability of finding a spacing
$E_t/\delta$ in the interval
$(s,s+ds]$\cite{RMT}.
   We see from Fig. \ref{Fig_1}(a) that the probability of finding a
spacing $E_t/\delta$ in the interval $(s,s+\Delta ]$
(histogram) is well described by
$P_W(s)\times p \Delta $ (thin-solid line) irrespective of the surface
criteria.

In general, given a set of $q$ levels, neighboring level distributions
$P_k(s)$ are defined as the probability of finding a spacing $E^k_t$
containing
$k$ adjacent levels in the interval $(s,s+ds]$. The case $k=0$
corresponds to the nearest level spacing distribution discussed above.
The statistical properties of $P_k(s)$  are well established
for  the GOE \cite{RMT}.
In particular, specific results hold for the standard deviation (or width)
$\sigma_k$ of these  distributions\cite{bohigas_annphys_1975}.
We now compare our numerical findings with
the GOE predictions.
Formally,
$(\sigma_k)^2=(\delta^{-2}/q)\sum^{q}_{t=1}(E^k_t-\delta_k)^2$
for the k-spacing: $E^k_t=\varepsilon_{k+t+1}-\varepsilon_{t}$ between
levels $|k+t+1\rangle$ and $|t\rangle$,
$\delta_k$ the mean of $E^k_s$, and $\delta=\delta_0$. Note that according
to this definition $\sigma_k$ is expressed in units of $\delta$.
Figure \ref{Fig_1}(b)  shows the GOE prediction (solid line) for
$\sigma_k$ and the values (symbols) found with the
above set of $p$ calculated energy levels for different disorder
realizations. For both surface criteria, the  calculated values of
$\sigma_k$ scatter around the GOE prediction.
For criterion ${\tt C_2}$, where
surface is a large fraction of the particle, the scatter is
stronger. Although the number of disorder realizations shown
is too small
for accurate ensemble averaging, the calculated
$\sigma_k$ are consistent with the RMT trend.


We now
consider the details of the electronic structure in a smaller region
around
$E_F$.
   Figure \ref{Fig_2} is a plot of the first few levels around
$|E_F\rangle$ for five disorder realizations for each surface criterion.
   The energy is measured relative to $E_F$  in units of $\delta^{Al}_{SP}$.
The levels present a rich structure and are far from
being equally-spaced by the energy $\delta^{Al}_{SP}$.  
We find that most of the levels form groups of two or more. The  results are
qualitatively similar   for ${\tt C_1}$  and
${\tt C_2}$, however, the details around
$E_F$ are set by disorder and surface criteria.
Note that the predicted number of levels ($\sim 19$) for all disorder
realizations (with exception to $\#1$ and $\#6$) studied here is fairly
consistent with the number of levels expected from the SP model in the
considered energy range: $[-10,10]\delta^{Al}_{SP}$.
    It is important to note that the {\em
clustering} seen in the spectra is taking place {\em in the presence of
disorder} in the metal-oxide interface. Although we have shown that in a
wider range of energy around $E_F$ the probability of finding strong 
bunching of
levels   is small, this {\em does not} mean that pairs or larger clusters
are avoided in a small neighborhood around $E_F$.


We now turn to the implications of this
clustering of energy levels for experimental tunnel resonance
spectra of the nanoislands.
   The orthodox  model of the device (nanoisland+tunnel junctions+leads)
   treats the leads as reservoirs of electrons with
chemical potentials
$\mu_L$ and $\mu_R$, coupled capacitively (with capacitances ${\cal C}_L$
and ${\cal C}_R$) to the nanoisland through the
   tunneling junctions. The applied voltage $V$ between the leads is
divided between     ${\cal C}_L$ and ${\cal C}_R$ setting the
    the electrochemical potential in the right(left) lead to
$\mu^{+(-)}=\mu_{R(L)}+(-)({\cal C}_{L(R)}/{\cal C}_{\Sigma})\,eV$ with
${\cal C}_{\Sigma}={\cal C}_{L}+{\cal C}_{R}$. When $V=0$,
$\mu_{L}$ and $\mu_{R}$
   are considered aligned with  $E_F$. As discussed by Averin and
Korotkov\cite{averin_JETP_1990}, the discrete energy spectrum of the
nanoisland allows tunneling to occur whenever
$\mu^{+(-)}$ (measured from $E_F$) matches the energy of one of the
available levels of the nanoisland.   Figure \ref{Fig_3} shows the
predicted \cite{comment_2}
$dI/dV$ spectra for
different disorder realizations and surface criteria.
The resonances are marked with flat-top(bottom)  segments indicating that
    $\mu^{+(-)}$ has originated the resonance.
    Two different regimes: i) (Fig. \ref{Fig_3}(a)) symmetric (${\cal
C}_L={\cal C}_R$), and ii) (Fig. \ref{Fig_3}(b)) non-symmetric  (${\cal
C}_L=1.5aF$ and ${\cal C}_R=3.2aF$\cite{ralph_experiments}) tunneling
junctions were considered.   The
$dI/dV$ spectra  arising from a model with equally spaced energy
levels is shown (solid circles) for comparison.  Clearly, the clustering
in the  energy spectra creeps into the $dI/dV$ spectra in a non-trivial
way. The nature of the clusters of resonances in the
$dI/dV$ spectrum is strongly dependent on disorder and on the relative
values of ${\cal C}_{L}$ and ${\cal C}_{R}$.

Now we show that the clustering of resonaces present in
the calculated $dI/dV$ qualitatively resembles the observed clustering. 
Experimentally (Fig. 1(a) of Ref. \cite{agam_PRL_1999}), the {\em
clusters} of resonances observed are as follows: i) a single resonance at the
begining of the spectrum (this sets the origin of energies), ii) 4 near-by
resonances around
$\delta^{Al}_{SP}$, iii) 2 near-by resonances located at
$1.5\delta^{Al}_{SP}$, and iv) 4 resonces distributed over an energy interval
equal to $\delta^{Al}_{SP}$  centered around
$2.5\delta^{Al}_{SP}$.
We  choose spectrum \#4 in the symmetric and non-symmetric regime,
under criterion $\tt{C_1}$, for comparison with the experiment.  
In the symmetric regime the spectrum shows: i) a
single resonance at the origin, ii) 2 pairs of resonances located around
$\delta^{Al}_{SP}$ and $2\delta^{Al}_{SP}$, respectively, and iii) 
4 near-by resonances spanning an energy interval approximately equal to
$\delta^{Al}_{SP}$, around
$3.5\delta^{Al}_{SP}$. When switching to the non-symmetric regime (the
experimental regime) the structure of the clustering is slightly modified,
with the overall number of resonces smaller. We see, then, that the observed
and calculated clusters show qualitative common features.  Whether the energy
scales ($\delta^{Al}_{SP}$) on which the theoretical and experimental
clusterings occur are similar or not remains an open question: This depends on
what the correct value of the effective volume of the {\em Al} island is. We
have shown that this volume may be significantly larger than has been
assumed previously. If this is the case then the experimental and
theoretical energy scales may be very
similar; letting the experiment be explained in a new way.

Finally, if we further assume that non-equilibrium
many-body effects such as those described by Agam {\em et
al.}\cite{agam_PRL_1999} are present then
still more clustering appears in the spectrum as
most of the single-electron
transport  resonances in Fig. \ref{Fig_3}
split to become clusters of resonances. However,
our results indicate
that the observed tunneling resonance spectra
should show a different behavior
from what would follow from the
simple equidistant-level model of the underlying
single-electron spectrum
that has been assumed in previous analyses
of the experiments, whether many-body non-equilibrium effects
are present or not.

%
%
%
%

Support from NSERC and CIAR is acknowledged.

%
%

\end{multicols}

%
%
\begin{table}

   \begin{tabular}{cccccccccccc}
     &  \multicolumn{5}{c}{${\tt C_1}$}   & &
\multicolumn{5}{c}{${\tt C_2}$}   \\ \cline{2-6}\cline{8-12}
label & $N_{O}$ & $N_{Al}$ & $n $ &  $E_F$ (eV) & $\delta $ (meV)
      & label  & $N_{O}$ & $N_{Al}$ & $n$ & $E_F$ (eV)  & $\delta $ (meV)  \\
\hline
$\# 1$ & 444 & 461 & 9093 & 8.297 & 2.34 &$\# 6$ & 651 & 623 & 9714 & 8.483 &
2.32
\\
$\# 2$ & 448 & 457 & 9105 & 8.313 & 2.30 &$\# 7$ & 634 & 640 & 9663 & 8.370 &
2.30
\\
$\# 3$ & 451 & 454 & 9114 & 8.330 & 2.30 &$\# 8$& 632 & 642 & 9657 & 8.383 &
2.32
\\
$\# 4$ & 440 & 465 & 9081 & 8.293 & 2.31 &$\# 9$& 630 & 644 & 9651 & 8.348 &
2.34
\\
$\# 5$ & 434 & 471 & 9063 & 8.251  & 2.32 &$\# 10$ & 629 & 645 & 9648 & 8.336 &
2.35
\\
\end{tabular}

\caption{Parameters characterizing the nanoparticles.
$C_1$ and $C_2$ are criteria that determine the surface (see text).}
\label{Table_1}

\end{table}


%
%
\begin{figure}
\caption{{\bf {\tt (a)}} Probability distribution (histograms) for
the nearest level spacings and Wigner
distribution (thin solid line).  {\bf {\tt (b)}} Standard
deviation
$\sigma_k$ for different disorder realizations (points) and the
GOE prediction (solid line). ${\tt C_1}$ and
${\tt C_2}$  as in Table 1.}
\label{Fig_1}
\end{figure}

%
%
\begin{figure}
\caption{Energy levels for different disorder realizations (see
text).
${\tt C_1}$ and ${\tt C_2}$ as in Table 1.}
\label{Fig_2}
\end{figure}

%
%
\begin{figure}
\caption{$dI/dV$ spectra for different disorder realizations
and capacitance regimes (see text).
${\tt C_1}$ and
${\tt C_2}$  as in Table 1.}
\label{Fig_3}
\end{figure}


\begin{references}
\bibitem{comment_2} Here we omit the Coulomb blockade energy $E_C$
of adding one electron to the nanoisland. Including this would cause
an upwards rigid shift of the resonances by an amount $E_C$.
Experimentally, this energy is $\sim 0.5meV$.
\end{references}
\end{document}